\documentclass[aps,prl,twocolumn,showpacs,amsmath,amssymb,preprintnumbers]{revtex4}

\newcommand{\vev}[1]{\langle#1\rangle}
\newcommand{\vect}{\left ( \begin{array}{c}}
\newcommand{\evect}{\end{array} \right )}

\usepackage{graphicx}
\usepackage{dcolumn}
\usepackage{bm}
\def\fsl#1{\setbox0=\hbox{$#1$}                 
   \dimen0=\wd0                                 
   \setbox1=\hbox{/} \dimen1=\wd1               
   \ifdim\dimen0>\dimen1                        
      \rlap{\hbox to \dimen0{\hfil/\hfil}}      
      #1                                        
   \else                                        
      \rlap{\hbox to \dimen1{\hfil$#1$\hfil}}   
      /                                         
   \fi}                                         %

\begin{document}

\preprint{\vtop{\hbox{RU06-4-B}\hbox{TKYNT-06-7}
\vskip28pt}}

\title{Higgs instability  in gapless superfluidity/superconductivity}

\author{Ioannis Giannakis$^1$, Defu Hou$^2$, Mei Huang$^{3,4}$,  Hai-cang Ren$^{1,2}$}

\affiliation{$^1$ Physics Department, The Rockefeller University, 1230 York Avenue, New York, NY 10021-6399}
\affiliation{$^2$ Institute of Particle Physics, Huazhong Normal University,Wuhan 430079, China}
\affiliation{$^3$ Physics Department, University of Tokyo, Hongo, Bunkyo-ku, Tokyo 113-0033, Japan}

\affiliation{$^4$ Institute of High Energy Physics, Chinese Academy of Sciences, Beijing 100039, China}

\begin{abstract}
In this letter we explore
the Higgs instability in the gapless superfluid/superconducting phase.
This is in addition to the (chromo)magnetic instability 
that is related
to the fluctuations of the Nambu-Goldstone bosonic fields. 
While the latter may induce a single-plane-wave LOFF state, 
the Higgs instability favors spatial inhomogeneity
and cannot be removed without a long range force.
In the case of the g2SC state the Higgs instability can only be partially 
removed by the electric
Coulomb energy. But this does not exclude the possibility 
that it can be completely removed in other exotic states such as 
the gCFL state.
\end{abstract}



\maketitle

Superfluidity or superconductivity with mismatched Fermi
momenta appears in many systems such as the
charge neutral dense quark matter, the asymmetric
nuclear matter, and in imbalanced cold atomic gases.
The mismatch plays the role of breaking the 
Cooper pairing. But it is not understood
how a BCS superconductor
is destroyed as the mismatch is increased. 
It was proposed in the 1960s that the competition between 
the pair breaking and the pair condensation 
would induce an unconventional superconducting phase, the
Larkin-Ovchinnikov-Fulde-Ferrel 
(LOFF) state \cite{LOFF-orig}.  Recently it was found that 
if the mismatch is larger than 
the gap magnitude, a gapless superconducting phase \cite{g2SC, gCFL}
or a breached pairing (BP) \cite{cold-atom-BP} can be formed in a charge 
neutral quark matter or in a constrained imbalanced 
cold atom system, respectively. Another scenario for the 
ground state at moderate mismatch is the phase separation\cite{mixed-phase}
between of superfluid/superconducting and normal phases.

Both gapless superconducting 
and BP phases exhibit instabilities. The former exhibits a
chromomagnetic instability 
\cite{chromo-ins-g2SC,chromo-ins-gCFL} while the latter a
superfluid density instability \cite{cold-atom-Wu-Yip}. Recent studies
have demonstrated that this type of instability is related 
to the local phase fluctuation of the superconducting
order parameter, i.e., the Nambu-Goldstone boson fields 
\cite{NG-current-hong, NG-current-huang, NG-Hashimoto, NG-current-gCFL}. 
This type of instability induces the formation of the single-plane-wave FF-like state 
\cite{FF-Ren, GHM-gluon, colored-FF, gCFL-LOFF, cold-atom-LOFF}. 

The imbalanced cold atom systems such as the $^{40}{\rm K}$
and $^{6}{\rm Li}$ offer an intriguing experimental opportunity to understand the 
pair breaking states. However, recent experiments \cite{cold-atom-Exp} on 
these systems did not show explicit evidence of  
the BP state or the LOFF state, rather they produced
strong evidence of the phase separation at moderate mismatch.
We explain in this letter that the BP state or the FF state maybe prevented 
to form there because of the Higgs instability induced by the 
mismatch. The Higgs instability is related to the magnitude fluctuation of the superconducting 
order parameter,  i.e., the Higgs field. It causes spatial inhomogeneity 
that may lead to phase separation. The Higgs instability was also considered in 
\cite{Kei-Kenji}, where it was named "amplitude instability".

The Higgs instability is an inhomogeneous extension of the Sarma instability against a 
homogeneous variation of the order parameter. Consider a system described by the 
Hamiltonian ${\cal H}$ with a spontaneous symmetry breaking, its thermodynamic 
potential density is given by 
\begin{equation}
\Omega=-\lim_{V\to\infty}\frac{T}{V}\ln{\rm Tr}e^{-\beta{\cal H}+\nu N},
\label{omega}
\end{equation}
where $N$ is a conserved charge with $\nu$ the corresponding chemical potential and $V$ 
is the volume of the system. The trace extends to a subset of the Hilbert space specified by 
an order parameter $\phi$, which is invariant under the broken symmetry group. The 
equilibrium condition reads $\left({\partial\Omega}/{\partial\phi}\right)_\nu=0$.
A homogeneous variation of $\phi$ around the equilibrium leads to $\delta\Omega
=\frac{1}{2}\left({\partial^2\Omega}/{\partial\phi^2}\right)_\nu\delta\phi^2+...$ \ \ .
The Sarma instability
\begin{equation}
\left(\frac{\partial^2\Omega}{\partial\phi^2}\right)_\nu<0
\end{equation}
renders the equilibrium unstable. In an ensemble of a fixed density of the $N$-charge,
\begin{equation}
n=-\left(\frac{\partial\Omega}{\partial\nu}\right)_\phi,
\label{constr}
\end{equation} 
however, the thermodynamic quantity to 
be minimized is the Legendre transformed free energy density, ${\cal F}=\Omega+\nu n$.
While the equilibrium condition $\left({\partial{\cal F}}/{\partial\phi}\right)_n=0$
is equivalent to $\left({\partial\Omega}/{\partial\phi}\right)_\nu=0$, 
the second order derivative reads
\begin{equation}
\Big(\frac{\partial^2{\cal F}}{\partial\phi^2}\Big)_n 
= \Big(\frac{\partial^2\Omega}{\partial\phi^2}\Big)_{\nu}
+\frac{\Big(\frac{\partial n}{\partial\phi}\Big)_{\nu}^2}
{\Big(\frac{\partial n}{\partial\nu}\Big)_{\phi}}.
\label{free}
\end{equation}
Because $\left({\partial n}/{\partial\nu}\right)_\phi=-\left({\partial^2\Omega}/
{\partial\nu^2}\right)_\phi>0$ as can be verified from the definition (\ref{omega}), 
$\left({\partial^2{\cal F}}/{\partial\phi^2}\right)_n
>\left({\partial^2\Omega}/{\partial\phi^2}\right)_\nu$. 
The removal of the Sarma instability by the constraint (\ref{constr}) amounts to 
$\left({\partial^2{\cal F}}/{\partial\phi^2}\right)_n>0$, leaving the 2nd order 
variation of the free energy
\begin{equation}
\delta{\cal F}=\frac{1}{2}\left(\frac{\partial^2{\cal F}}{\partial\phi^2}\right)_n\delta\phi^2>0
\label{variation}
\end{equation}
at the equilibrium. This is, however, {\it{not enough}}. 
The inclusion of inhomogeneous variations 
$\phi_{\vec k}$ extends (\ref{variation}) to
\begin{equation}
\delta{\cal F}=\frac{1}{2}\left(\frac{\partial^2{\cal F}}{\partial\phi^2}\right)_n\delta\phi^2
+\frac{1}{2}\sum_{\vec k\neq 0}\left(\frac{\partial^2{\cal F}}
{\partial\phi_{\vec k}^*\partial\phi_{\vec k}}\right)_n
\delta\phi_{\vec k}^*\delta\phi_{\vec k},
\end{equation}
with $\left({\partial^2{\cal F}}/{\partial\phi_{\vec k}^*\partial\phi_{\vec k}}\right)_n
=\left({\partial^2\Omega}/{\partial\phi_{\vec k}^*\partial\phi_{\vec k}}\right)_{\nu}$. 
As long as
\begin{equation}
\lim_{k\to 0}\lim_{V\to\infty}
\left(\frac{\partial^2\Omega}{\partial\phi_{\vec k}^*\partial\phi_{\vec k}}\right)_\nu
=\left(\frac{\partial^2\Omega}{\partial\phi^2}\right)_\nu<0,
\label{variationk}
\end{equation}
the Sarma instability will return with respect to $\delta\phi_{\vec k}$ of sufficiently 
low but nonzero $k$ without offsetting the globally imposed constraint (\ref{constr}).  

The Higgs instability can also be understood intuitively. Let us divide the
whole system being considered into subsystems. As long as the size of each
subsystem is much larger than all interaction length scales, its
thermodynamics is identical to that of the master system but with
the constraint relaxed. The Sarma instability will develop through
different variations of the order parameters of each subsystem while
maintaining the constraint conditions of the master system.

To analyze the Higgs instability of the g2SC,  
we start from the bosonized 2-flavor Nambu--Jona-Lasinio (NJL) model,
the Lagrangian density has the form of
\begin{eqnarray}
\label{lagr2}
{\cal L}_{2SC} & =  & {\bar q}(i\fsl{\partial}+\hat \mu \gamma^0)q 
-\frac{1}{4G_D} \Delta^{*\rho}\Delta^{\rho}
 \nonumber \\
 & + & \frac{i\Delta^{*\rho}}{2}[\bar q^Ci\gamma^5\tau_2\epsilon^{\rho} q]
-\frac{i\Delta^{\rho}}{2}[\bar q i\gamma^5\tau_2 \epsilon^{\rho}q^C].
 \label{lagr-2sc}
\end{eqnarray} 
The requirement of $\beta$-equilibrium induces the mismatch between 
the Fermi surfaces of the pairing quarks.  The matrix of chemical potentials 
in the color-flavor space ${\hat \mu}$ is given in terms of the quark number 
chemical 
potential $\mu$, the color chemical potential  $\mu_8$, and the electric 
chemical potential $\mu_e\equiv 2 \delta\mu$. 
Because $\mu_8 \simeq O(\Delta^2/\mu)<<\Delta$, 
we safely take $\mu_8 =0$ in our weak coupling calculations. 
Then $\mu_u= \bar\mu-\delta\mu$ and $\mu_d =\bar\mu+\delta\mu$ with 
$\bar\mu=\mu-\delta\mu/3$. The conserved charge $N$ and its chemical potential 
$\nu$ of the general discussion above correspond to the electric charge $Q$ and the electric chemical potential $\mu_e$ of the model.
The $\rho$th anti-triplet composite diquark field reads
$\Delta^{\rho}=2iG_D(\bar q^Ci\gamma^5\tau_2 \epsilon^{\rho}q)$, where $\epsilon^{\rho}$ is 
an anti-symmetric matrix in the color space 
with $\rho=1,2,3$ indicating respectively 
the anti-red, anti-green, and anti-blue colors of the diquark.
In the 2SC phase, the color symmetry $SU(3)_c$ is spontaneously broken to $SU(2)_c$
and diquark field obtains a nonzero expectation value. 
Without loss of generality, one can always assume that diquark condenses in the
anti-blue direction, the ground state of the 2SC phase is 
characterized then by  $\vev{\Delta^{3}}\equiv \Delta>0$, and $\vev{\Delta^{1}}=
\vev{\Delta^{2}}=0$.  

The most general spatial fluctuation of the order parameter can be parameterized as
\begin{equation}
\vect\Delta^1(\vec r)\\ \Delta^2(\vec r)\\ \Delta^3(\vec r)\evect =
\exp \left[ i \sum_{a=4}^8 \varphi_a(\vec r) T_a  \right] \vect0\\0\\ \Delta + H(\vec r) \evect,
\label{nl-field}
\end{equation}
where $\varphi_a$'s are five Nambu-Goldstone bosons \cite{NG-current-huang}, 
$T_a$'s are the corresponding generator of $SU(3)_c$ and $H$ the real Higgs field. 
The Fourier components of $H$ correspond to $\delta\phi$ and $\delta\phi_{\vec k}$
of Eqs. (\ref{variation})-(\ref{variationk}).
For the rest of the letter, we focus only on the Higgs mode by setting $\varphi_a=0$,
since $H$ and $\varphi$'s do not mix quadratically. 

The thermodynamic potential of the system including Higgs fluctuation reads
$\Omega = \Omega_M + \Omega_{H}$.
The mean-field free-energy at equilibrium is given by
\begin{eqnarray}
\Omega_M= - \frac{T}{2} \sum_n\int\frac{d^3\vec{p}}{(2\pi)^3} {\rm Tr} \ln ( [{\cal S}_M(P)]^{-1}) + \frac{\Delta^2}{4 G_D},
\end{eqnarray}
where the inverse propagator ${\cal S}_M^{-1}$ takes the form of
\begin{equation}
\left[{\cal S}_M(P)\right]^{-1} = 
\gamma_0(p_0+\bar\mu\rho_3)-\vec\gamma\cdot\vec p-\delta\mu\gamma_0\tau_3\rho_3-\Delta\gamma_5\epsilon^3\tau_2\rho_2,
\label{prop-0}
\end{equation}
with $\rho$'s the Pauli matrices with respect to NG indexes
and the four momentum $P=(p_0,\vec p)$.
The g2SC state is characterized by $\Delta<\delta\mu$.
The explicit form of $\Omega_M$ has been given in Ref.~\cite{g2SC} with 
the Sarma instability \cite{Sarma},
\begin{equation}
\left(\frac{\partial^2\Omega}{\partial\Delta^2}\right)_{\mu_e} 
= \frac{4\bar\mu^2}{\pi^2}\Big[1-\frac{\delta\mu}{\sqrt{(\delta\mu)^2-\Delta^2}}\Big]<0
\label{sarma-g2SC}
\end{equation}
in weak coupling, i.e. $\delta\mu<<\bar\mu$.  

The thermodynatic potential of the Higgs field reads
\begin{equation} 
\label{free-energy-H}
\Omega_{H} = \frac{1}{2}\int\frac{d^3\vec k}{(2\pi)^3}H^*(\vec k)\Pi(k) H(\vec k),
\end{equation}
with the self-energy function
\begin{eqnarray}
\Pi(k) &= & \frac{1}{4 G_D} 
 - \frac{T}{4} \sum_n\int\frac{d^3\vec{p}}{(2\pi)^3}  \nonumber \\
& & {\rm tr} \Big [ 
{\cal S}_M (P)\gamma_5\epsilon^3\tau_2\rho_2
 {\cal S}_M(P-K)\gamma_5\epsilon^3\tau_2\rho_2\Big],
\end{eqnarray}
where the trace extends to Dirac, color, flavor and NG indexes and $K=(0,\vec k)$. 
At $T=0$, we obtain the result of $\Pi(k)=A_H+B_Hk^2$ for $k<<\Delta$
with $A_H=\left({\partial^2\Omega}/{\partial\Delta^2}\right)_{\mu_e}$ given in 
Eq. (\ref{sarma-g2SC}) and
\begin{equation}
B_H=\frac{2\bar\mu^2}{9\pi^2\Delta^2}\Big[1-\frac{(\delta\mu)^3}
{((\delta\mu)^2-\Delta^2)^{\frac{3}{2}}}
\Big].
\label{BH}
\end{equation}
The negative $A_H$ signals the Higgs instability and the negative $B_H$ indicates that the 
instability gets stronger at nonzero $k$. Note that our coefficient $B_H$ is more negative than that of
Ref. \cite{Kei-Kenji}. 
Numerically, we found that the function $\Pi(k)$ reaches a minimum at some intermediate $k$ before 
increasing with $k$. For $k>>\Delta$, we find $\Pi(k)\simeq(\bar\mu^2/\pi^2)(\ln(k/\Delta)
+{\rm const.})>0$. 

Unless there exists a competing mechanism that results in a positive contribution to the Higgs 
self-energy function $\Pi(k)$ for $k\neq 0$. The Higgs instability will prevent the exotic states of 
gapless superfluidity/superconductivity from being implemented in nature. 
This explains the failure of observing BP state in cold neutral atoms. In case 
of quark matter, however, the long range Coulomb interaction of the electric charge 
fluctuation induced by the inhomogeneous Higgs field has to be examined  
before reaching a conclusion. 

\begin{figure}
\begin{minipage}[t]{0.442\textwidth}
\includegraphics[width=8cm]{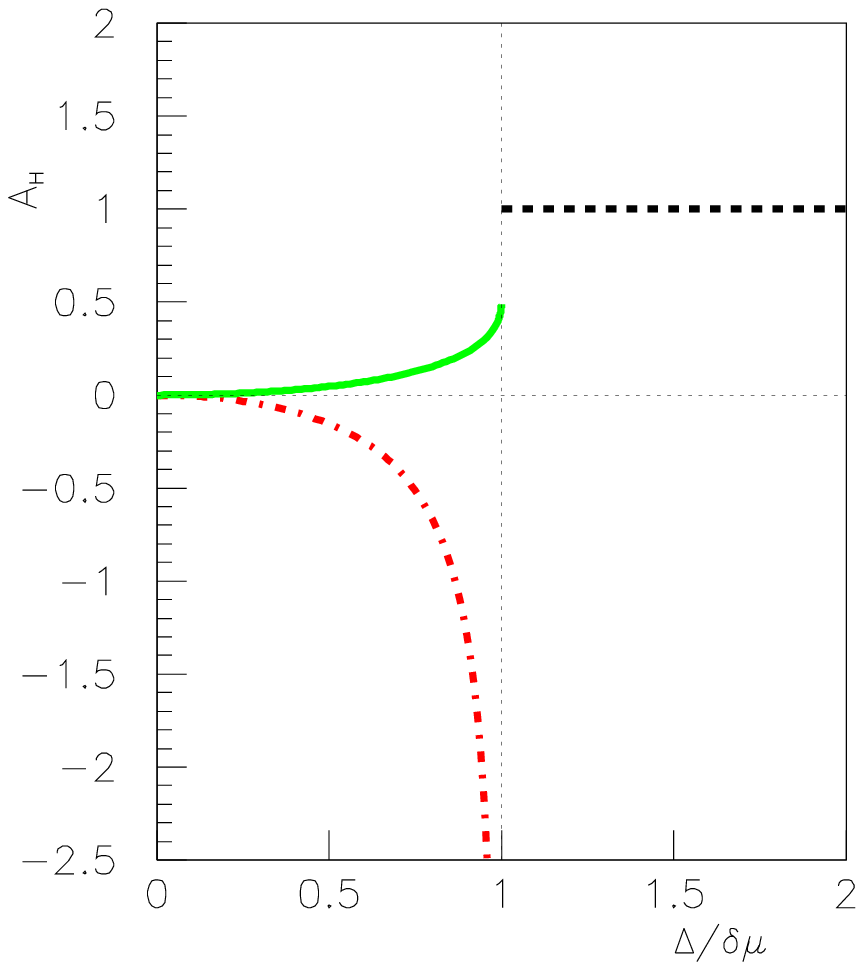}
\caption{\label{Higgs-f-fig}
$A_H$(red dashed-dotted line) and $\tilde A_H$ (green solid line)
are plotted as functions of $\Delta/\delta\mu$ in the g2SC phase, and the
black dashed line shows the corresponding quantity in the 2SC phase.}
\end{minipage}
\hspace{0.1\textwidth}
\begin{minipage}[t]{0.442\textwidth}
\includegraphics[width=8cm]{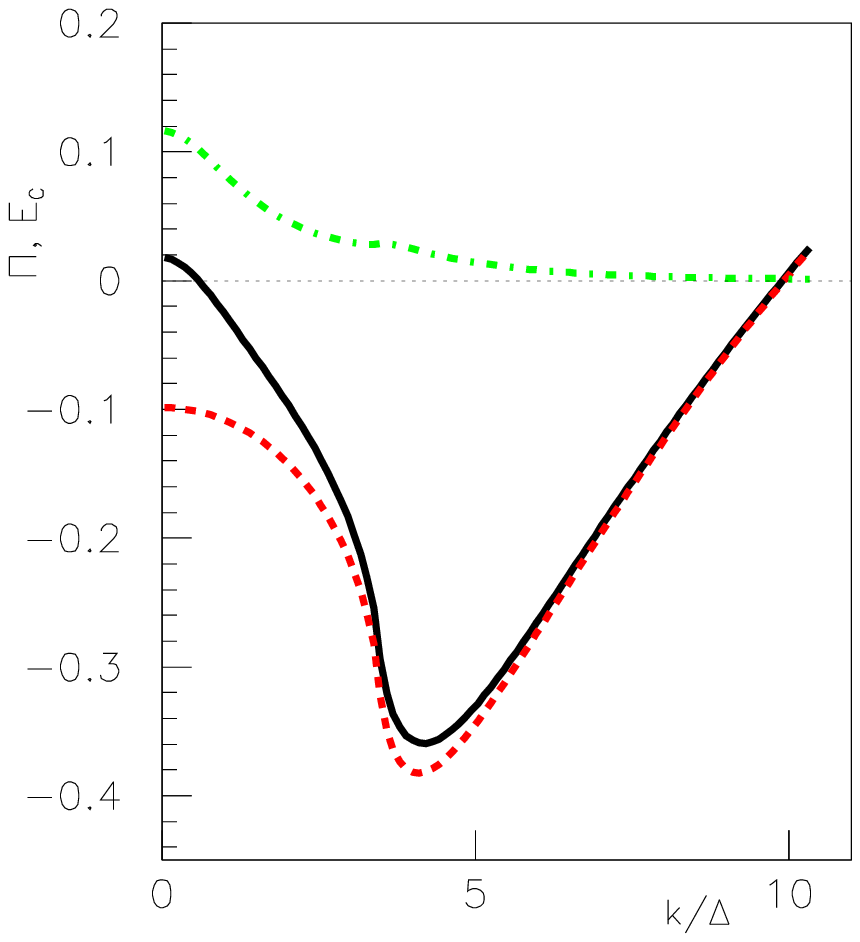}
\caption{\label{Higgs-k-fig}
The function $\Pi(k)$ (red dashed line), $\tilde\Pi(k)$ (black solid line) and the Coulomb
energy (green dash-dotted line) as functions of scaled-momentum $k/\Delta$,  in the
case of $\delta\mu=2\Delta$ and $(e^2\bar\mu^2)/(4\pi\Delta^2)=1$.}
\end{minipage}
\end{figure}

The inhomogeneity generated Coulomb energy density that should be added to RHS of
$\Omega_H$ of (\ref{free-energy-H}) reads
\begin{equation}
E_{\rm coul.}=\frac{1}{2V}\sum_{\vec k\neq 0}\frac{\delta\rho(\vec k)^*\delta\rho(\vec k)}
{k^2+m_D^2(k)}
\end{equation} 
where $\delta\rho(\vec k)$ is the Fourier component of the Higgs-induced charge density
and $m_D$ is the Coulomb polarization function ( Debye mass at $\vec k=0$ ). We have 
$\delta\rho(\vec k)=\kappa(k)H(\vec k)$, with
\begin{equation}
\kappa(k)=-\frac{eT}{2V}\sum_P{\rm tr}\gamma_0Q{\cal S}_M(P)
\gamma_5\epsilon^3\tau_2\rho_2{\cal S}_M(P-K),
\end{equation}
and
\begin{equation}
m_D^2(k)=-\frac{e^2T}{2V}\sum_P{\rm tr}\gamma_0Q{\cal S}_M(P)
\gamma_0Q {\cal S}_M(P-K),
\end{equation}
where the charge matrix $Q=\rho_3(a+b\tau_3)$ with $a=1/6$ and $b=1/2$ 
for the quark matter consisting of $u$ and $d$ flavors.  
It can be shown that the Goldstone fields do not generate charge fluctuations.
The Legendre transformed free energy of Higgs field is obtained from 
(\ref{free-energy-H}) with $\Pi(k)$ replaced by  
\begin{equation}
\tilde\Pi(k)\equiv
\Big(\frac{\partial^2{\cal F}}{\partial H^*(\vec k)\partial H(\vec k)}\Big)_{n_Q}
=\Pi(k)+\frac{\kappa^*(k)\kappa(k)}{k^2+m_D^2(k)}.
\label{coulomb}
\end{equation}
Notice that 
$m_D^2(0)=e^2({\partial n_Q}/{\partial\mu_e})_{\Delta}$
and $\kappa(0)=e({\partial n_Q}/{\partial\Delta})_{\mu_e}$ with $n_Q$ the 
charge density.
The second term on RHS of (\ref{free}) is included in the second term of 
(\ref{coulomb}) in the infinite volume limit, invalidating the equality 
(\ref{variationk}). We have
\begin{eqnarray}
\kappa(0)&=&\frac{4eb\bar\mu^2}{\pi^2}\frac{\Delta}{\sqrt{\delta\mu^2-\Delta^2}},  \\
m_D^2(0)&=&\frac{2e^2b^2\bar\mu^2}{\pi^2}\Big(1+\frac{2\delta\mu}{\sqrt{\delta\mu^2-\Delta^2}}\Big).
\end{eqnarray}
On writing $\tilde A_H\equiv\tilde\Pi(0)$, we obtain that
\begin{equation}
\tilde A_H=\frac{4(b^2-3a^2)\bar\mu^2
(\delta\mu-\sqrt{\delta\mu^2-\Delta^2})}{\pi^2[3a^2\sqrt{\delta\mu^2-\Delta^2}
+b^2(2\delta\mu+\sqrt{\delta\mu^2-\Delta^2})]}>0
\end{equation}
for the whole range of g2SC state. This quantity together with that without Coulomb term are 
plotted in Fig.\ref{Higgs-f-fig}.

The detailed calculation of the form factor $\kappa(k)$, function $m_D(k)$
as well as the Coulomb improved Higgs self-energy in the whole mementum $k$ 
space will be reported in another paper \cite{long-paper}. 
We find that there is always some domain of $k$ where $\tilde\Pi(k)<0$ 
throughout the entire region of g2SC state ($\delta\mu>\Delta$) for
an arbitrary ratio $m_D(0)/\Delta$. Therefore the electric Coulomb energy is 
not strong enough to cure the Higgs instability of g2SC for all momenta. 
See Fig. \ref{Higgs-k-fig} for an example.  It is noticed that $\Pi(k)$ reaches
its minimum at a rather large momentum, i.e., $k \simeq 4 \Delta$, which
indicates the formation of phase separation \cite{Igor}.

In this letter, we have explored the Higgs instability 
against local fluctuations
of the magnitude of the order parameter that are not prohibited
by global constraints.
In the case of the imbalanced neutral atom systems, the instability
extends from arbitrarily low momenta-albeit not zero-to momenta
much higher than the inverse coherence length. Without a long range force 
a phase separation is likely to be the structure of the ground state since it minimizes 
the gradient energy of the inhomogeneity. 
For the case of two flavor quark matter that is globally neutral, the instability
can be removed by the electric Coulomb energy for low and high momenta.
But it remains in a window of intermediate momenta.
Nevertheless, the long range Coulomb interactions might favor
a ground state with crystalline structure. Possible candidates
include the heterotic mixture of normal and superconducting phases
\cite{mixed-phase} and the multi-plane-wave LOFF state \cite{lo-phase}. 
Although our calculations are
limited to the case of g2SC, the existence of the Higgs instability
in the absence of Coulomb interactions is universal for all
gapless superfluidity/superconductivity that are subject to the
Sarma instability. Our general formulation
from (\ref{omega}) to (\ref{variationk}) can be trivially generalized to the system
with several invariant order parameters and constraints. It would
be extremely interesting to examine whether the electric and color
Coulomb energies are capable of eliminating the Higgs instability completely
in the gCFL phase.

\vskip 0.3 cm

\noindent{\bf Acknowledgments:}
We thank M. Forbes, K. Fukushima, K. Iida, M. Hashimoto, T. Hatsuda, D.K. Hong, 
I. Shovokovy and  P.F. Zhuang for stimulating discussions. The work of I.G. and H.C.R 
is supported in part by US Department of Energy 
under grants DE-FG02-91ER40651-TASKB. The work of D.F.H. is supported in part 
by Educational Committee under grants NCET-05-0675 and 704035. The work of D.F.H 
and H.C.R. is also supported in part by NSFC under grant No. 10575043.  The work of 
M.H. is supported in part by the Japan Society for the Promotion of Science Fellowship 
Program and the Institute of High Energy Physics, Chinese Academy of Sciences, she 
also would like to thank the APCTP for its hospitality.

\end{document}